\documentclass[%
 amsmath,amssymb,
 aps,prl,reprint,
]{revtex4-2}

\usepackage{graphicx}    
\usepackage{subcaption}  
\usepackage{orcidlink}   
\usepackage{booktabs}
\usepackage{comment}
\hypersetup{
    colorlinks=true,
    linkcolor=blue,
    filecolor=blue,
    urlcolor=blue,
    citecolor=blue
}
\usepackage{ragged2e}

\begin{document}

\preprint{APS/123-QED}

\title{Quantum Kinetics of Fast-Electron Inelastic Collisions in Partially-Ionized Plasmas}

\author{Yeongsun Lee\orcidlink{0000-0003-4474-416X}}
 \affiliation{Department of Nuclear Engineering, Seoul National University, Seoul, South Korea \\
 Nuclear Research Institute for Future Technology and Policy, Seoul National University, Seoul, South Korea}

\author{Pavel Aleynikov\orcidlink{0009-0002-3037-3679}}
 \affiliation{Max-Planck Institute fur Plasmaphysik, Greifswald, Germany}

\author{Jong-Kyu Park\orcidlink{0000-0003-2419-8667}}
\email{jkpark@snu.ac.kr}
\thanks{Corresponding author.}
\affiliation{%
 Department of Nuclear Engineering, Seoul National University, Seoul, South Korea
 }%

\date{\today}

\begin{abstract}
Fast electrons in partially ionized plasmas lose energy through inelastic collisions with bound electrons. While the mean energy loss is well described by stopping-power theory, fluctuations associated with discrete excitation and ionization events produce energy straggling and an additional longitudinal diffusion in momentum space. We incorporate this effect into fast-electron kinetics through a derived Fokker–Planck operator whose coefficients are obtained from \textit{ab initio} quantum many-body simulations. We demonstrate that neglecting inelastic energy diffusion in partially ionized D–Ar plasmas can underestimate primary runaway-electron generation by several orders of magnitude.
\end{abstract}

\maketitle
\textit{Introduction.}--
Inelastic interactions between fast electrons and atoms are among the fundamental processes in nature. To understand how fast electrons lose energy in matter, Bethe’s theory of stopping power \cite{Bethe1932} has been widely applied across various fields \cite{Andersen1989PRL, Gurevich1992PLA, Scheidenberger1994PRL, Baro1995NIM, Zylstra2015PRL, Hesslow2017PRL} with further refinements provided by modern quantum mechanics \cite{Salvat2022PRA}. However, the stopping power formulation does not describe the small deviations in the energy loss experienced by fast electrons during ionization losses \cite{Williams1929PRSA}. These small deviations result in the so-called energy straggling \cite{Williams1929PRSA, Landau1944, Landau1944c, Fano1963AR}, such that an initially mono-energetic beam is not only slowed down, but its energy distribution also acquires a finite width. In a case of energetic particles passing through a thin layer, this asymmetric distribution is often called the Landau energy loss distribution \cite{Landau1944, Landau1944c}.

\begin{figure}
    \centering
    \includegraphics[width=0.8\linewidth]{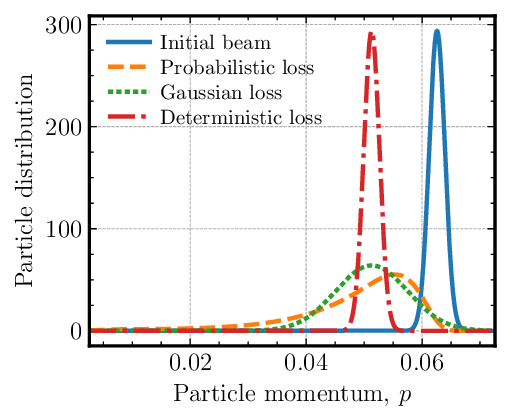}
    \caption{\justifying
    Particle density distribution of an electron beam undergoing collisional slowing down from initial energy $1\,\mathrm{keV}$ (blue solid) in neutral Ne gas ($n_{Ne}=10^{20}\,\mathrm{m^{-3}}$). For inelastic energy loss: red dashed dotted---convective slowing-down only; green dotted---convective-diffusive (Gaussian) approximation; orange dashed---probabilistic energy transfers for $\Delta t = 0.66  \, \mathrm{\mu s}$, corresponding to about five inelastic collision times. Initial beam width is $5 \, \%$ of the final energy spreading.
    }
    \label{fig:slowing-down}
\end{figure}

This phenomenon appears difficult to reconcile with classical kinetic theory in plasmas. In long-range Coulomb interactions, energy diffusion does not occur against stationary targets (at the binary-collision level) \cite{Rosenbluth1957PR} and only arises when many-body effects are properly incorporated through the non-stationary distribution of the field particles \cite{Helander2005Cam}. Meanwhile, when a purely deterministic stopping-force model is applied to binary encounters between electrons and neutrals, the stochastic broadening of the energy distribution (energy straggling) does not appear. Indeed, the red curve in Fig. \ref{fig:slowing-down} shows the convective slowing down of an electron beam with the initial energy $1\,\mathrm{keV}$ and finite width (blue) in phase space, while an \textit{ab initio} quantum many-body simulation \cite{Sun2018WIR, Sun2020JCP} suggests that the probabilistic energy loss should produce a spread in particle energy (orange curve). Evidently, a quantum kinetic description is necessary to account for the energy loss straggling from inelastic collisions beyond the classical theory as it requires solving the quantum many-body problem within the atom at every binary encounter.

A critical problem for which quantum kinetics can offer a resolution is the accurate description of primary runaway electron generation in partially-ionized plasmas \cite{Wilson1925MPCPS, Dreicer1959PR, Connor1975NF, Lee2024PRL, Chiu1998NF, Harvey2000POP, Gurevich1961JETP, Smith2008PoP, Aleynikov2017NF, Svenningsson2021PRL}. The consideration of inelastic energy straggling was originally addressed as the energy diffusion by Gurevich in a pioneering work \cite{Gurevich1961JETP}, but has since been largely overlooked in modern runaway-kinetic studies of atmospheric physics \cite{Roussel1994PRE, Dwyer2014PR}, inertial fusion \cite{Mosher1975PoF,Robinson2014NF} and magnetized fusion \cite{Martin-solis2017NF, Hesslow2019JPP}. Even the dedicated studies that further developed the quantum description \cite{Walkowiak2022PoP, Savoye-Peysson2023NF, Walkowiak2025ADNDT} primarily considered elastic scattering and the mean excitation energy.

One limitation of the deterministic description is that no electric-field–driven upward particle flux can form near the mean force-balance momentum in neutral-dominated plasmas. Suppose that the application of an electric field shifts this momentum, i.e. runaway momentum, to $0.06$ in Fig. \ref{fig:slowing-down}. In this case most electrons represented by the red curve will decelerate. However, particles in the distribution shown by the orange curve that experience weaker-than-average friction, i.e. $\mathrm{p>0.06}$, can in fact still be accelerated by the electric field and form an upward particle flux through the energy diffusion.

This Letter shows that accounting for inelastic energy diffusion is essential for accurately interpreting runaway-electron generation during disruption experiments in present-day tokamaks. In a representative DIII-D disruption scenario \cite{Hollmann2016NF}, for instance, the relevant plasma parameters ($n_D=2\times10^{19}\,\mathrm{m^{-3}}$, $T_e=5\,\mathrm{eV}$, $E=90\,\mathrm{V\,m^{-1}}$, and $n_{Ar}=3.31\times10^{19}\,\mathrm{m^{-3}}$) give a characteristic inelastic energy-diffusion frequency of $\sim3.91\times10^4\,\mathrm{s^{-1}}$. This is comparable to the Coulomb energy-diffusion frequency ($\sim3.38\times10^4\,\mathrm{s^{-1}}$). Because runaway generation depends exponentially on the balance between electric-field acceleration and collisional drag, as well as the collisionality of energy diffusion, this additional diffusion can increase the predicted primary runaway-electron generation by several orders of magnitude.

\textit{Quantum kinetic theory.}--
Within the first Born approximation, assuming non-relativistic bound electrons but allowing a fast projectile electron the differential cross section of inelastic scattering is given by \cite{Landau2013}
\begin{equation}
    d\sigma_n = \frac{d\sigma_{R}}{dq} |F_{n0}(\vec{q})|^2dq, \label{eq:dsigman_Ruther_unnorm}
\end{equation}
where $F_{n0}(\vec{q}) \equiv \sum_{a=1}^Z \int e^{-i \vec{q} \cdot \vec{r}_a} \psi_n^* \psi_0 d\tau$ is the inelastic (transition) form factor. $d\tau=dV_1 \cdots dV_Z$ is the differential volume element of the Z orbital electrons, $\vec{p} = \hbar\vec{k}$ and $\vec{p} ' =\hbar \vec{k}'$ are the relativistic momenta of an incident electron before and after inelastic collision, respectively, and $\vec{q} \equiv \vec{k}' - \vec{k}$ is the recoil variable. $\frac{d\sigma_R}{dq} \equiv 8 \pi \Big( \frac{e^2}{4\pi \varepsilon_0\hbar v} \Big)^2 \frac{1}{q^3}$ is the Rutherford cross section. When $n=0$, $F_{n0}(\vec{q})$ becomes the elastic form factor $F_{00}(\vec{q})$.

If inelastic scattering is dominated by small energy transfers, the short-term evolution of a mono-energetic electron beam can be described by the Landau distribution \cite{Landau1944, Landau1944c}. Yet, in a confined system, the electrons remain in plasma and can undergo many successive inelastic interactions with a partially ionized background. These interactions cumulatively drive the long-term evolution of their distribution function toward a Gaussian core with a low energy tail arising from rare high-energy-loss events, as discussed by Bohr \cite{Bohr1948}. Therefore, the Gaussian approximation is not exact, but it can be usefully adopted when (1) the background plasmas vary on a timescale much longer than the inelastic collision time, (2) the particle energy greatly exceeds the mean inelastic energy loss and (3) the dynamics of interest is controlled primarily by the core of the distribution and are insensitive to the effects of large energy-loss events (knock-ons). In this limit, the Fokker–Planck (FP) collision operator \cite{Rosenbluth1957PR} adequately describes the resulting collision dynamics:
\begin{equation}
\begin{split}
    \mathcal{C}^{in}[{f_e}]  &= - \vec{\nabla}_p \cdot (f_e \Delta \vec{p}) + \frac{1}{2} \vec{\nabla}_p\vec{\nabla}_p: (f_e \Delta \vec{p} \Delta \vec{p}). \label{eq:FP_expanded}
\end{split}
\end{equation}
Let $E_0$ be the energy of atom in the ground state and let $E_n$ be the energy of the  $n$-th state after the collision. Energy conservation then gives $\frac{p^2 - p'^2}{2m \gamma} \approx E_n - E_0$, where we have used $\gamma + \gamma' \approx 2 \gamma$, which is valid in the Born-approximation regime of interest, $q \ll k$. When electrons in a single phase space cell encounter many atoms within a unit flight time, we can evaluate the expectation values $\Delta\vec{p}=-\sum_I\mathcal{A}^I\hat{p}$ and $\Delta\vec{p}\Delta\vec{p}=\sum_I\mathcal{D}^I_{p\perp} (\overleftrightarrow{I} - \hat{p}\hat{p}) + \mathcal{D}_{p||}^I\hat{p}\hat{p}$  without loss of generality. Here $I$ labels the species of neutral and partially ionized atoms and
\begin{equation}
    \mathcal{A}^I = n_I v \hbar\int \sum_n \Big(\frac{q^2}{2k} +\frac{m\gamma}{k\hbar^2} (E_n-E_0)\Big) d\sigma_n \label{eq:A},
\end{equation}
\begin{equation}
    \mathcal{D}^I_{p\perp} = \frac{n_I v \hbar^2}{2} \int \sum_n \Big(q^2 - \Big(\frac{q^2}{2k} + \frac{m\gamma}{k\hbar^2} (E_n-E_0) \Big)^2\Big) d\sigma_n \label{eq:D_perp}
\end{equation}
and
\begin{equation}
    \mathcal{D}^I_{p||} = n_I v \hbar^2 \int \sum_n \Big(\frac{q^2}{2k} + \frac{m\gamma}{k\hbar^2} (E_n-E_0) \Big)^2 d\sigma_n. \label{eq:D_para}
\end{equation}
Note that $\mathcal{D}^I_{p||}$ is typically neglected in modern fusion-related runaway kinetic studies \cite{Mosher1975PoF, Robinson2014NF,Martinsolis2015PoP,Hesslow2017PRL,Hesslow2018JPP}.

The range of integration is divided into two parts, from $q_{\min}$ to $q_0$ and from $q_0$ to $q_{\max}$, respectively, where $v_0/v \ll q_0 a_0 \ll 1$ and $v_0/v \ll q_{\max} a_0 \ll v/v_0$. Depending on a value of $q$, we can separate the $n$-summation \footnote{This is replaced with integral for ionizing collisions to consider continuum spectrum.} and $q$-integral. For a high-$q$, the Bethe sum rule simplifies the $n$-summation and subsequently facilitates the $q$-integral. The maximum value of the recoil variable is $q_{\max}\equiv \frac{p}{\sqrt{2} \hbar}$ \cite{Bethe1997}. For small $q$, the dipole approximation $F_{0n}\approx - i q d_{0n}$, where $d_{0n}\equiv \sum_{a=1}^Z \int x_a \psi_n^* \psi_0 d\tau$, facilitates the $q$-integral and subsequently the Thomas-Reiche-Kuhn (TRK) sum rule simplifies the $n$-summation. The minimum value of the recoil variable is $q_{\min}\equiv \frac{E_n-E_0}{\hbar v}$. 
\begin{table}[h!]
    \centering
    \begin{tabular}{c|c|c}
        \toprule
        $\kappa$ & $\sum_{n\geq1} ( E_n - E_0)^\kappa |F_{n0}|^2$ & $\sum_{n\geq1} ( E_n - E_0)^\kappa |d_{n0}|^2$ \\
        \midrule
        0 & $\displaystyle S$ & $\displaystyle (d^2_x)_{00}$ \\
        1 & $\displaystyle \frac{Z\hbar^2 q^2}{2m}$  & $\displaystyle \frac{Z\hbar^2}{2m}$ \\
        2 & $\displaystyle \frac{2ZT_0\hbar^2q^2}{3 m} + \frac{ S\hbar^4 q^4}{4m^2}$ & $\displaystyle \frac{2ZT_0\hbar^2  }{3 m}$ \\
        \bottomrule
    \end{tabular}
    \caption{\justifying
    Generalized Bethe sum rules ($|F_{0n}|^2$) and TRK sum rules ($|d_{0n}|^2$) for different $\kappa$.
    }
    \label{tab:sum_rules}
\end{table}

A list of sum rules \cite{Fano1963AR, Fano1968RMP} is summarized in Table \ref{tab:sum_rules}, where $T_0\equiv \frac{(|\sum_{a=1}^Z \vec{p}_a|^2)_{00}}{2mZ}$ \footnote{The first moment of the TRK sum rule is often denoted as $S(1) \equiv \Sigma_n(E_n-E_0) N_{n0}$. Here we instead define $T_0 \equiv \frac{3S(1)}{4Z}$ and $T_b \equiv \frac{S(1)}{2Z}$ for convenience in the kinetic description.} and $S=Z - \sum_{i=1}^Z \sum_{j=1}^Z (e^{i \vec{q}\cdot(\vec{r}_j - \vec{r}_i)})_{00}$. We retain the correlation effect between particles ($i\neq j$) through $(\vec{p}_{i} \cdot \vec{p}_{j})_{00}$ in $T_0$ and $(e^{i \vec{q}\cdot(\vec{r}_j - \vec{r}_i)})_{00}$ in $S$ \cite{Heisenberg1927}. When $(\vec{p}_{i} \cdot \vec{p}_{j})_{00}$ is negligible, $T_0$ reduces to a convenient measure of the average kinetic energy of bound electrons in the ground state.

Incorporating the sum rules into the integrals in Eq. \ref{eq:A}, $\mathcal{A}^I$ can be written in a form equivalent to the Bethe stopping-power expression
\begin{equation}
    \mathcal{A}^I \approx \frac{n_IZe^4}{4\pi \varepsilon_0^2pv} \Big( \gamma\ln \frac{\hbar v q_{\max}}{I} + \ln \frac{q_{\max}}{q_s} + \mathcal{O}(q_0a_0)\Big),
\end{equation}
where $I$ is the mean excitation energy, defined by a weighted sum $\ln I \equiv \frac{1}{Z} \sum_{n} N_{n0}\ln(E_n - E_0) $ with the oscillation strength $N_{n0} \equiv \frac{2m}{\hbar^2}(E_n-E_0)|d_{n0}|^2$ as suggested by Bethe. The logarithmic term $Z\ln \frac{q_{\max}}{q_s}$ effectively approximates $\int_{q_0}^{q_{\max}}\frac{S}{q}dq$, since the dominant contribution of the integral lies well above $q_0$ \cite{Heim2000JPB}. From Eq. \ref{eq:D_perp}, $\mathcal{D}^I_{p\perp}$ can be written
\begin{equation}
    \mathcal{D}^I_{p\perp} \approx \frac{n_IZe^4}{4\pi \varepsilon_0^2v} \Big( \ln \frac{q_{\max}}{q_s} - \frac{2\gamma^2 mT_0}{3p^2} \ln \frac{\hbar v q_{\max}}{I_1} +  \mathcal{O}(q_0a_0)\Big),
\end{equation}
where the inequality $\int_{q_0}^{q_{\max}}Zqdq \gg \int_{q_0}^{q_{\max}}Sqdq$ has been used. Because pitch-angle scattering is dominated by elastic (coherent) processes $\sim Z^2$ rather than inelastic (incoherent) processes $\sim Z$, $\mathcal{D}^I_{p\perp}$ is expected to play a relatively minor role in the overall electron kinetics, consistent with the relation $q_s \gg \frac{I}{\hbar v}$. From Eq. \ref{eq:D_para}, $\mathcal{D}^I_{p||}$ becomes
\begin{equation}
    \mathcal{D}^I_{p||} \approx \frac{n_IZe^4}{2\pi \varepsilon_0^2v} \Big(\frac{2\gamma^2 mT_0}{3p^2} \ln \frac{\hbar v q_{\max}}{I_1} +  \mathcal{O}(q_0a_0)\Big).
\end{equation}
$I_1$ is the straggling mean excitation energy, defined as $\ln I_1 \equiv \frac{3}{4 T_0} \sum_{n} (E_n-E_0) N_{0n} \ln(E_n - E_0)$ \cite{Fano1963AR}.

We normalize $p\to \tilde{p}mc$, $v \to \beta c$, $T_0 \to \tilde{T}_0mc^2$. After suppressing the tilde for brevity and defining the quasi-temperature of bound electrons as $T_b \equiv \frac{2}{3} T_0$, in analogy with the free electron temperature definition, Equation \ref{eq:FP_expanded} can be written in terms of characteristic frequencies,
\begin{equation}
    \mathcal{C}^{in}[{f_e}] =  \frac{1}{p^2} \frac{\partial}{\partial p} \Big(p^3 ( \nu_S + \frac{1}{2}p \nu_{||}\frac{\partial}{\partial p}) \Big) f_e + \nu_D \mathcal{L}( f_e) \label{eq:linearFP}
\end{equation}
where the slowing down frequency $\nu_S$ is
\begin{equation}
    \nu_S = \sum_I 4\pi c r_e^2 \frac{Zn_I \gamma^2}{p^3} \Big(\ln \frac{\hbar v q_{\max}}{I} + \frac{3T_b}{\gamma} \ln \frac{\hbar v q_{\max}}{I_1} \Big)  \label{freq:slowing},
\end{equation}
the deflection frequency $\nu_D$ is
\begin{equation}
    \nu_D = \sum_I 4\pi c r_e^2 \frac{Zn_I \gamma}{p^3}\Big( \ln \frac{q_{\max}}{q_s} - \frac{\gamma^2T_b}{p^2} \ln \frac{\hbar v q_{\max}}{I_1} \Big),
\end{equation}
and the energy diffusion frequency $\nu_{||}$ is
\begin{equation}
    \nu_{||}= \sum_I 8\pi c r_e^2 \frac{Zn_I \gamma^3}{p^5} T_b \ln \frac{\hbar v q_{\max}}{I_1} \label{freq:para}.
\end{equation}
$\mathcal{L}\equiv \frac{1}{2} \frac{\partial}{\partial \mu} (1-\mu^2) \frac{\partial}{\partial \mu}$ is the pitch-angle scattering operator, $\mu$ is cosine of the pitch angle, and $r_e \equiv \frac{e^2}{4\pi\varepsilon_0 mc^2}$ is the classical electron radius. These frequencies introduce the longitudinal diffusion term $\nu_{||}$, which represents the dominant new effect, while the finite-$T_b$ contributions to $\nu_S$ and $\nu_D$ appear only as small corrections to the deterministic friction model \cite{Robinson2014NF,Martinsolis2015PoP,Hesslow2017PRL,Hesslow2018JPP}.

\begin{figure}
    \centering
    \includegraphics[width=\linewidth]{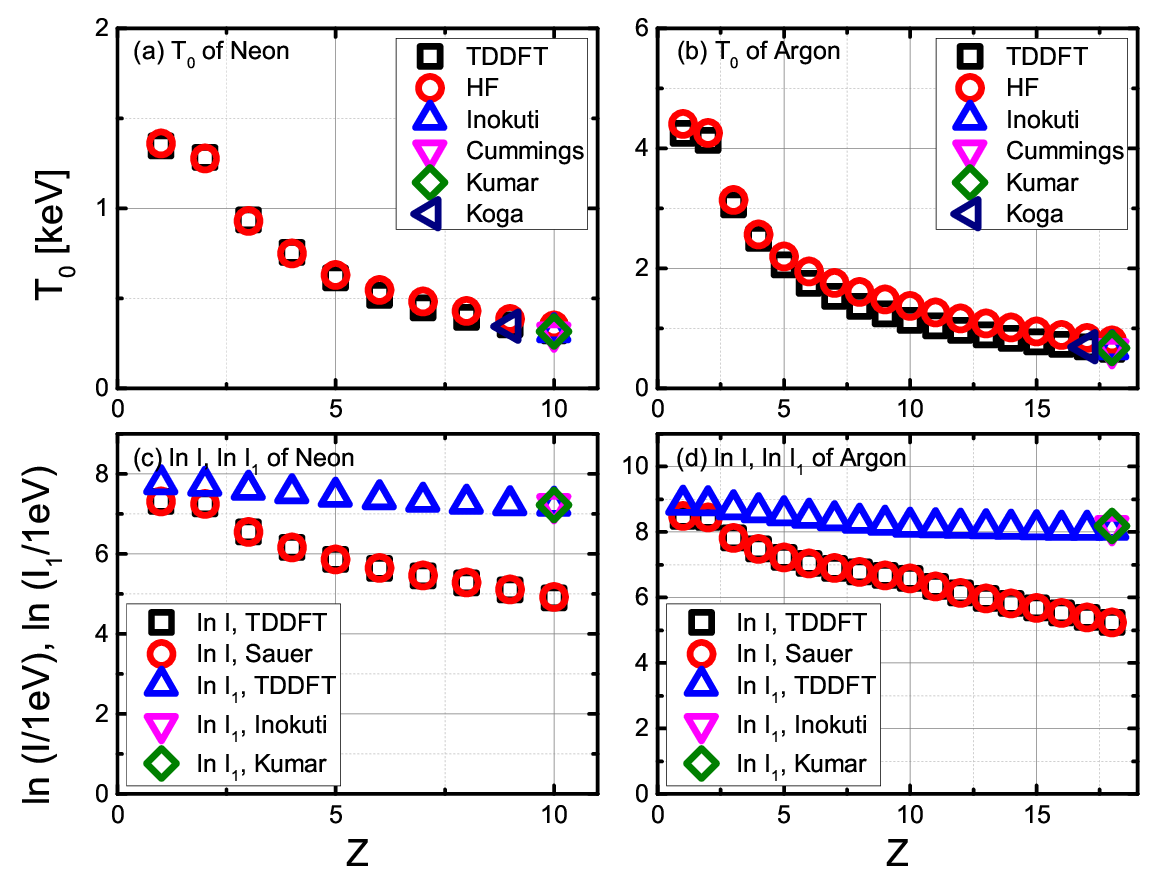}
    \caption{\justifying
    The TDDFT results of (a,b) $T_0$ in keV, (c,d) $\ln I$ and $\ln I_1$ in eV for Ne (a,c) and Ar (b,d), respectively. The HF calculation in (a,b) is the averaged kinetic energy of orbital electrons. Reference results, identified by author names in the legend, are included for comparison \cite{Inokuti1978PRA,Cummings1975JCP,Kumar2010JCP, Koga2002JCP,Sauer2015AQC,Sauer2018JCP}.
    }
    \label{fig:tddft}
\end{figure}
The use of three parameters ($T_0$, $I$, $I_1$), rather than the single $I$ of Bethe's theory, allows for a compact representation of inelastic processes through Eq. \ref{eq:linearFP}. This generalization, however, requires determining $T_0$ and $I_1$ of partially stripped atoms in addition to $I$. While the simple Thomas–Fermi (TF) model \cite{Thomas1927MPCPS, Fermi1928ZP, Kirillov1975} is sufficient to describe the elastic deflection frequency \cite{Breizman2019NF}, accurate quantum calculations are required because inner-shell contributions to $T_0$ and $I_1$ are significant \cite{Inokuti1978PRA}. Hence, we perform time-dependent density functional theory (TDDFT) calculations for neutral and ionized states of Ne and Ar using the PySCF package \cite{Sun2018WIR, Sun2020JCP} with the aug-cc-pCV5Z basis set \cite{Dunning1989JCP} and the CAM-B3LYP functional \cite{Yanai2004JPL}. Ne and Ar are selected because of their relevance to the critical problem of runaway generation in fusion plasmas discussed later. Figures \ref{fig:tddft}(a,b) show the resulting $T_0$. Our TDDFT results for neutrals and singly charged states of Ne and Ar agree well with previous calculations \cite{Cummings1975JCP, Inokuti1978PRA, Koga2002JCP, Kumar2010JCP}. They also converge to the mean kinetic energy given by the Hartree–Fock (HF) model \cite{Hartree1928MPCPS, Fock1930ZP} in the hydrogen-like atomic states, i.e. Ne$^{+9}$ and Ar$^{+17}$, where correlation effects vanish. Figures \ref{fig:tddft}(c,d) show $\ln I$ and $\ln I_1$. The accuracy in constructing the oscillation strength $N_{n0}$ is validated by the agreement of $\ln I$ with Dalton package's results in Refs. \cite{Sauer2015AQC,Sauer2018JCP}. Our $I_1$ of Ne and Ar are also consistent with Refs. \cite{Inokuti1978PRA, Kumar2010JCP}. For ionized states, reference data are largely unavailable. Instead, we verified our methodology by benchmarking against ionized Al \cite{Oddershede1990NIM} although those results are not shown here.

Effective interaction parameters $Z$, $T_b$, $I$, $I_1$ depend on electron energy thus, rigorously speaking, Eqns (\ref{freq:slowing}, \ref{freq:para}) are only valid for highly energetic electrons. In fact, more than $69 \, \%$ and $74 \, \%$ of $T_b$ are accounted for by energies greater than $1 \, \mathrm{keV}$ in Ne and Ar, respectively \cite{Kumar2010JCP}. To extend applicability to mildly relativistic electrons, we modify $\nu_S$ and $\nu_\|$ by restricting the sum rules to states satisfying $E_n - E_0 \leq mc^2\frac{\beta\gamma}{\sqrt{2}}$ and replacing $Z$, $T_b$, $I$, $I_1$ with effective parameters $Z^{\text{eff}}$, $T_b^\text{eff}$, $I^\text{eff}$, $I_1^\text{eff}$ (e.g. see Figs. 1-2 in Ref. \cite{Kamakura2006JAP}). Note that the Born approximation formally requires $\beta \gtrsim Z^\text{eff}\alpha$, where $\alpha$ is the fine-structure constant. For the mildly relativistic regime considered here this condition is only marginally satisfied, although partial screening by inner-shell electrons effectively reduces the interaction strength and extends the practical range of validity.

\begin{figure}
    \centering
    \includegraphics[width=\linewidth]{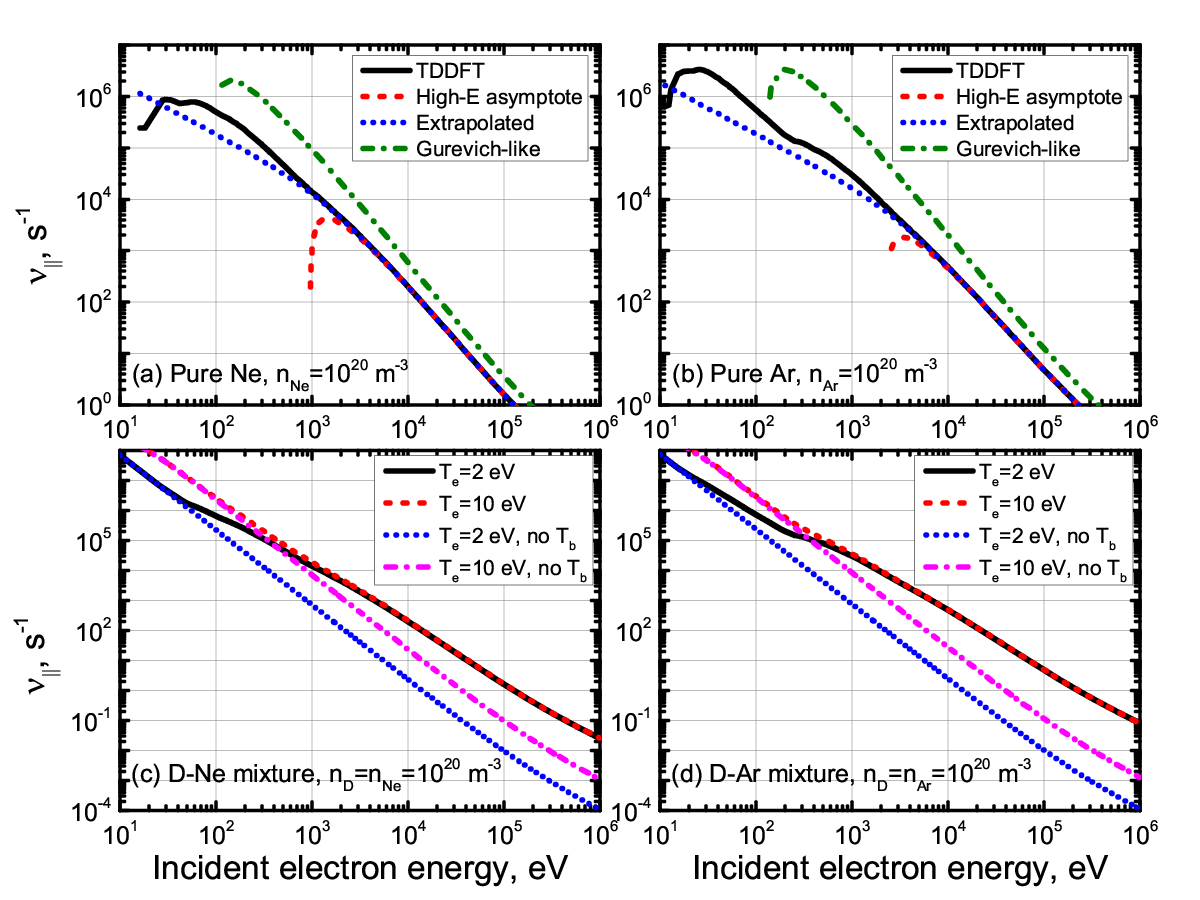}
    \caption{\justifying
    The net energy diffusion frequency $\nu_\parallel$ in pure impurity gases (a,b) and mixture plasmas (c,d). Impurity species are Ne for (a,c) and Ar for (b,d).
    }
    \label{fig:nup}
\end{figure}

Figure \ref{fig:nup} demonstrates $\nu_\|$ as a function of particle kinetic energy. In Figs. \ref{fig:nup}(a,b), we consider neutral gases, where the energy diffusion is entirely due to inelastic collisions. The black curve shows $\nu_\|$ evaluated with the effective parameters and agrees well with the high energy asymptote given by Eqn. \ref{freq:para} (red dashed curve). We also show that a simple extrapolating expression, replacing $\ln \frac{\hbar v q_{\max}}{I_1}$ with $\frac{1}{1.5}\ln (1+(\frac{\hbar v q_{\max}}{I_1})^{1.5})$ provides a practical approximation of $\nu_\|$ above $1 \, \mathrm{keV}$.

Note that a related longitudinal diffusion coefficient $\nu_\|$ was considered by Gurevich for neutral \cite{Gurevich1961JETP}. That treatment was formulated within a deterministic stopping-force description and therefore relied on several simplifying assumptions. When expressed within the present quantum-mechanical formulation, several refinements emerge: (1) the numerical coefficient in Eq. \ref{freq:para} is $8\pi$, a factor of 1.5 smaller than $12\pi$ obtained by applying the virial relation to Eq. (15a) of Ref. \cite{Gurevich1961JETP}, (2) the correlation effect $(\vec{p}_{i} \cdot \vec{p}_{j})_{00}$ in $T_0$ is retained, (3) the straggling mean excitation energy in treated separately from ionisation potential, $\ln I_1 \neq\ln I$ and (4) the energy dependence of the effective atomic parameters is explicitly included. To illustrate the effect of these microscopic resolutions, we plot a Gurevich-type approximation (green dash-dotted curve) defined as $\sum_I 12\pi c r_e^2 \frac{Zn_I \gamma^3}{p^5} T_b \ln \frac{\hbar v q_{\max}}{I}$ using TDDFT values of $T_b$ and $I$. Within the present kinetic framework this approximation produces a substantially larger estimate of $\nu_\|$, whereas the refined formulation yields a correspondingly smaller value.

For $\nu_\|$ in mixture plasmas, we show two representative cases in which the free electron temperatures $T_e$ is either $2$ or $10 \, \mathrm{eV}$ in Figs. \ref{fig:nup}(c,d). Coulomb energy diffusion is included using Equation (3.47) of Ref. \cite{Helander2005Cam}. At these temperatures deuterium is fully ionized, whereas impurities are partially ionized and their charge balances are described by the collisional radiative equilibrium (CRE) model \cite{Summers2004ADAS}. In the high-energy region, the critical role of inelastic collisions in energy diffusion is clearly visible through a substantial reduction in the net $\nu_\|$ (blue, magenta) when bound-electron contribution is removed $T_b = 0$. Although the atomic charge states differ between the $T_e=2 \, \mathrm{eV}$ (black) and $10 \, \mathrm{eV}$ (red) cases, $\nu_\|$ remains almost unchanged, reflecting the similar contributions from the inner-shell electrons to the straggling integral.
\begin{table}[h!]
    \centering
    \begin{tabular}{@{} c|c|c @{}}
        \toprule
        & Coulomb collisions & Inelastic collisions \\
        \midrule
        $\displaystyle \nu_S$ & $\displaystyle 4\pi n_ec r_e^2 \frac{ \gamma^2}{p^3} \ln \frac{b_{max}}{b_{min}}$ & $\displaystyle 4\pi n_b c r_e^2 \frac{\gamma^2}{p^3} \ln \frac{\hbar v q_{\max}}{I}$ \\
        $\nu_D$ & $\displaystyle 4\pi n_e c r_e^2 \frac{\gamma}{p^3}\ln \frac{b_{max}}{b_{min}}$  & $\displaystyle 4\pi n_b c r_e^2 \frac{\gamma}{p^3}\ln \frac{q_{\max}}{q_s}$ \\
        $\displaystyle \nu_{||}$ & $\displaystyle 4\pi n_e c r_e^2 \frac{\gamma^3}{p^5} 2T_e \ln \frac{b_{max}}{b_{min}}$ & $\displaystyle 4\pi n_b c r_e^2 \frac{\gamma^3}{p^5} 2T_b \ln \frac{\hbar v q_{\max}}{I_1}$ \\
        \bottomrule
    \end{tabular}
    \caption{\justifying
    Similarity in the leading order part in characteristic frequencies between classical Coulomb collisions and quantum inelastic collisions. $b_{min}$ and $b_{max}$ are the minimum and maximum impact parameters, respectively.
    }
    \label{tab:freqs}
\end{table}

The collision operator for quantum inelastic interactions has a structure similar to that of classical Coulomb collisions, although the microscopic physics is different. In Coulomb kinetics plasma interactions can be decomposed into many weak binary encounters between charged particles.  In contrast, inelastic scattering involves a fast projectile interacting with a bound multi-electron system whose dynamics is intrinsically many-body. 
Despite this difference, the leading contributions to the characteristic frequencies can be written in terms of effective densities ($n_e$, $n_b$) and energy scales ($T_e$, $T_b$) of free and bound electrons, as summarized in Table \ref{tab:freqs}.

Incident energy dependence also arises in both problems but for different reasons. In Coulomb collisions it originates from collective plasma effects that modify the maximum interaction scale \cite{Bohm1949PR}. In inelastic collisions it instead reflects the atomic excitation spectrum: higher projectile energies allow access to higher electronic states and additional participating orbital electrons. Finally, in Coulomb collisions, a single Coulomb logarithm describes the range of impact parameters in binary encounters. In inelastic collisions, however, the logarithmic terms correspond to different moments of the atomic excitation spectrum, leading to the parameters $I$ and $I_1$ that characterize mean energy loss and energy straggling.

In Coulomb kinetics, electrons experience numerous interactions, naturally leading to a Gaussian distribution of energy loss by the central limit theorem. In contrast, a Gaussian approximation is not trivial for inelastic collisions, as discussed in the paragraph above Eq. \ref{eq:FP_expanded}. The energy-loss distribution is typically characterized as a observable function of the path length traversed by particles in a medium in detector physics \cite{Landau1944, Bohr1948, Fano1963AR}. In confined plasmas, however, the temporal evolution of the distribution in momentum space is of primary interest. This problem can be formulated as an analysis of the corresponding Green’s function.

Figure \ref{fig:slowing-down} demonstrates that the Gaussian (FP-like, green curve) Green’s function $G_{FP}$ reasonably approximates the Boltzmann-like Green’s function $G_{B}$ (orange curve) produced by successive inelastic collisions. The particle density distribution is obtained by the time evolution of the initial Dirac-delta distribution with the initial momentum $p_0$: The Gaussian Green's function is locally approximated as $G_{FP}(p,\Delta t)=\frac{1}{\sqrt{4\pi D \Delta t}}\exp(-\frac{(p-(p_0-u\Delta t))^2}{4D \Delta t})$ with $u=p(\nu_S - \nu_\|)$ and $D=\frac{1}{2}p^2 \nu_\|$ whereas $G_{B}$ is numerically constructed from the collision frequency involving inelastic energy transition to the $n$-th state, $n_I\sigma_n v = \frac{n_I e^4}{2\pi \varepsilon_0^2 \hbar^2 v} \int_{q_{\min}}^{q_{\max}} |F_{n0}|^2\frac{dq}{q^3}$ \footnote{The constructed $G_B$ from the TDDFT results underestimated the stopping power by $5.3 \%$, originating from the high-$q$ region.}. For the comprehensive illustration, we consider the initial Gaussian distribution with finite width (blue curve) and choose the visualizing cell size sufficiently large so as to simplify features arising from discrete energy loss. The slowing-down process alone (red curve) merely shifts the Dirac-delta function and therefore fails to properly capture the inelastic energy diffusion that originates from quantum-level uncertainty. As a result, it does miss electrons that have undergone few or no collisions.

While free electrons can gain energy via Coulomb energy diffusion during relaxation toward a Maxwellian distribution, inelastic collisions do not allow incident particles to gain energy. Formation of upward flow is possible in the presence of an electric field. However, the Gaussian tail can produce unphysical acceleration of some particles by an amount of 
\begin{equation}
    \int_{p\geq p_0} G_{FP}dp = \frac{1}{2}\text{erfc}(u\sqrt{\frac{\Delta t}{4D}}). \label{eq:accel_pop}
\end{equation}
In Fig. \ref{fig:slowing-down}, such a particle fraction is only about $3 \, \%$ \footnote{This fraction initially reaches $50 \, \%$ as $\Delta t \to 0$ and subsequently decreases with time. However, a comparison in this short-time regime is not meaningful because it simply reflects the breakdown of the Gaussian approximation before sufficient collisional accumulation. The estimated value of $3 \, \%$ should thus be interpreted as indicating the fraction of particles forming the unphysical upward flow remains small if the Gaussian approximation is valid.}.

\textit{Primary generation of runaway electrons.}--
In realistic applications to primary generation of runaway electrons, the FP description of inelastic collisions encounters regions of phase space where its validity is limited for two distinct reasons. First, when the particle kinetic energy is comparable to the ionization potential energy, inelastic energy loss does not result in an accumulated Gaussian distribution underlying Eq. \ref{eq:linearFP}. Second, the Taylor expansion of $f_e$ taken in Eq. \ref{eq:FP_expanded} treats momentum changes as small, wheres inelastic collisions produce finite momentum decrements, unlike the infinitesimal momentum variations of Coulomb collisions. Consequently, the FP formulation imposes a restrictive condition on the gradient of the distribution function. The description remains physical only if the inelastic collisions produce a downward diffusive flux, which requires
\begin{equation}
    \frac{\partial}{\partial p} \ln f_e \geq - \frac{p \nu_\|}{2 \nu_S}.\label{eq:cond_downward}
\end{equation} 
Nevertheless, the physically relevant range of solutions can be bounded by two limiting asymptotic descriptions, which approximate the behavior expected from the binary Boltzmann operator \cite{Lee2024PRL} for low energy electrons, where the FP approximation is no longer valid. The first is the "collisionless" (CL) asymptote, obtained by neglecting inelastic energy loss, i.e.
\begin{equation}
   \nu_\| \to \min (\nu_\|, -\frac{2 \nu_S}{p \frac{\partial}{\partial p} \ln f_e}). \label{eq:const_cl}
\end{equation}
The second is the "slowing-down" (SD) asymptote, obtained by neglecting the finite-$T_b$ correction,
\begin{equation}
    T_b \to 0. \label{eq:const_sd}
\end{equation}
These asymptotic limits are applied only in phase-space regions where their condition Eq. \ref{eq:cond_downward} is violated.
\begin{figure}
    \centering
    \includegraphics[width=\linewidth]{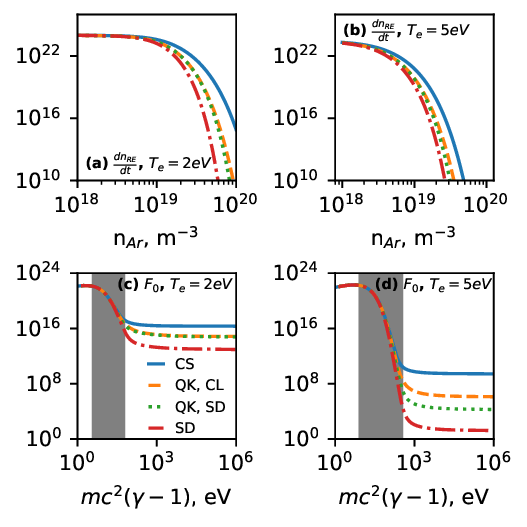}
    \caption{\justifying
    (a), (b) $\frac{dn_{RE}}{dt}$ in $m^{-3}s^{-1}$ as a function of $n_{Ar}$ with $n_D = 2\times10^{19}\,m^{-3}$. (c), (d) The corresponding distribution function $F_0$ yielding steady-state $\frac{dn_{RE}}{dt}$ when $n_{Ar} = 3.31\times 10^{19}\,\mathrm{m^{-3}}$. $T_e = 2 \, \mathrm{eV}, \, E=230 \, \mathrm{V \, m^{-1}}$ for (a), (c) and $T_e = 5 \, \mathrm{eV}, \, E=90 \, \mathrm{V\, m^{-1}}$ for (b), (d), respectively. Blue solid curve neglects free-bound inelastic collisions. Orange dashed and green dotted curves use Eq. \ref{eq:linearFP} with the CL and SD asymptotes, respectively. Red dashed dotted curve merely takes $T_b=0$ in all phase space regions. Gray-shaded areas show the region where the collisionless asymptote (Eq. \ref{eq:const_cl}) is applied.
    }
    \label{fig:fss}
\end{figure}

We implement the quantum-kinetic collision operator in the kinetic simulation \cite{Aleynikov2017NF} using the FiPy finite-volume partial differential equation solver \cite{Guyer2009CSE}. In this implementation, the coefficients  $\nu_S$ and $\nu_\|$ in Eq. \ref{eq:linearFP} are obtained from the TDDFT calculations shown in Fig. \ref{fig:nup}, while the deflection frequency $\nu_D$ is neglected.

In Fig. \ref{fig:fss}, the influence of of inelastic collisions on the so-called Dreicer generation \cite{Dreicer1959PR, Gurevich1961JETP, Connor1975NF, Lee2024PRL} is investigated in partially ionized D-Ar mixture plasmas. The plasma parameters correspond to conditions relevant to the early current quench phase in the DIII-D tokamak \cite{Hollmann2016NF}, where an electric field is inferred from the initial ohmic current density $j_0\approx 0.85 \, \mathrm{MA\, m^{-2}}$ using Ohm's law. The ionization states are determined assuming the CRE. The runaway electron generation rate $\frac{dn_{RE}}{dt}$ is largest when inelastic interactions are completely screened (CS) and only free-free collisions are considered, as indicated by the blue curves in Figs. \ref{fig:fss}(a-b). Inelastic energy loss reduces $\frac{dn_{RE}}{dt}$ (orange or green curves), and the result obtained with the CL asymptote consistently exceeds that with the SD asymptote. However, if inelastic energy diffusion is neglected ($T_b = 0$), friction-only treatment (red curve) underestimates $\frac{dn_{RE}}{dt}$ by up to several orders of magnitude. This sensitivity arises because the Dreicer generation rate depends exponentially on the effective energy diffusion near the critical momentum. Note that the remaining uncertainty in our model originates from the choice of the asymptotic treatments, which define upper (CL) and lower (SD) bounds on $\frac{dn_{RE}}{dt}$. Our estimate for $\frac{dn_{RE}}{dt}$ is therefore reliable particularly when the two bounds are close.

Figures \ref{fig:fss}(c-d) show the particle distribution functions $F_0 \equiv \int d\mu 2\pi p^2 f_e (p,\mu)$ after reaching a steady state when $n_{Ar}=3.31\times 10^{19}\mathrm{m^{-3}}$. Gray-shaded region indicated where the CL asymptote is applied \footnote{For the comparative one-dimensional illustration, the phase space region is determined using $F_0(p)$, not $f_e(p,\mu)$ by translating Eq. \ref{eq:cond_downward} to $\frac{\partial}{\partial p}\ln F_0 \geq - \frac{2(\nu_S - \nu_\|)}{p \nu_\|}$.}. In Fig. \ref{fig:fss}(c), $F_0$ from the CL and SD solutions are nearly identical and the corresponding generation rates are $\frac{dn_{RE}}{dt}$ are also close, namely $8.95 \times 10^{19} \, \mathrm{m^{-3} s^{-1}}$ and $7.92 \times 10^{19}\, \mathrm{m^{-3} \,s^{-1}}$, respectively. In contrast, in Fig. \ref{fig:fss}(d), $F_0$ shows different levels of plateau structure, yielding the distinct generation rates of $\frac{dn_{RE}}{dt}=6.74\times 10^{10}\, \mathrm{m^{-3} s^{-1}}$ and $1.00 \times 10^{9}\, m^{-3} s^{-1}$, respectively. The origin of this discrepancy is related to the sharpeness of the transition of $F_0$ forms from the Maxwellian core to the runaway region. In $T_e = 2 \, \mathrm{eV}$ plasmas (Figs. \ref{fig:fss}(a,c)), relative to $T_e = 5 \, \mathrm{eV}$ plasmas (Figs. \ref{fig:fss}(b,d)), the magnitude of $\frac{\partial}{\partial p}\ln f_e$ is smaller, because the electric field is stronger and weakly-ionized Ar produces few free electrons. In the above four cases, we estimated the contribution of the unphysical acceleration by inserting the simulation time step into Eq. \ref{eq:accel_pop} at the boundary cell adjacent to the gray regions in Figs. \ref{fig:fss}(c,d) with the same number of particles. The resulting unphysical contribution is less than $1\, \%$ of $\frac{dn_{RE}}{dt}$ and is therefore negligible.


\textit{Summary.}--
In summary, we extend the kinetic description of inelastic collisions of fast electrons in partially ionized plasmas by incorporating atomic coefficients from time-dependent density functional theory (TDDFT) into a Fokker–Planck framework. This formulation connects the classical stopping-power picture to the stochastic broadening of energy losses discussed by Landau and later treatments of straggling: while individual inelastic collisions produce discrete, one-sided energy losses, their cumulative effect can be represented by a Gaussian core with an associated longitudinal diffusion coefficient. We show that this inelastic energy diffusion plays an essential role in the primary generation of runaway electrons. Under conditions relevant to present-day tokamak disruptions, neglecting it can underestimate the primary runaway-electron generation rate by several orders of magnitude.

The authors are grateful to Prof. Per Helander for useful suggestions. This research was supported by R\&D Program of "Optimal Basic Design of DEMO Fusion Reactor, CN2502-1" through the Korea Institute of Fusion Energy (KFE) funded by the Government funds, and also by the Technology Development Projects for Leading Nuclear Fusion through the National Research Foundation of Korea (NRF) funded by the Ministry of Science and ICT (No. RS-2023-00281276).

\bibliographystyle{unsrt}
\bibliography{ref}

\end{document}